\UseRawInputEncoding

\documentclass[journal]{IEEEtran}  

\usepackage{multirow}
\usepackage[table,xcdraw]{xcolor}

\usepackage{amsmath,amsfonts}
\usepackage{algorithmic}
\usepackage{array}
\usepackage[caption=false,font=footnotesize,labelfont=rm,textfont=rm]{subfig}
\usepackage{textcomp}
\usepackage{stfloats}
\usepackage{url}
\usepackage{verbatim}
\usepackage{graphicx}

\usepackage{color}
\usepackage{amssymb}
\usepackage{booktabs}
\usepackage[table]{xcolor}
\usepackage{graphicx}
\usepackage{multirow}
\usepackage{color} 
\usepackage{cite}
\usepackage{mathrsfs}

\newcommand{\figref}[1]{Fig. \ref{#1}}

\newcommand{\ra}[1]{\renewcommand{\arraystretch}{#1}}

\newcommand{\tabref}[1]{Table \ref{#1}}

\usepackage{algorithm}
\usepackage{algorithmic}

\newcommand{\alref}[1]{Algorithm \ref{#1}}

\def\BibTeX{{\rm B\kern-.05em{\sc i\kern-.025em b}\kern-.08em
		T\kern-.1667em\lower.7ex\hbox{E}\kern-.125emX}}
\usepackage{balance}
\usepackage{hyperref}
\hypersetup{hidelinks} 
\hypersetup{
	    colorlinks=true,
		linkcolor=blue
	}

\begin{document}
	\title{EnvCDiff: Joint Refinement of Environmental Information and Channel Fingerprints via Conditional Generative Diffusion Model}

\author{
	Zhenzhou~Jin,~\IEEEmembership{Graduate Student Member,~IEEE,}
	Li~You,~\IEEEmembership{Senior Member,~IEEE,}\\
	Xiang-Gen~Xia,~\IEEEmembership{Fellow,~IEEE,}
	and~Xiqi~Gao,~\IEEEmembership{Fellow,~IEEE}
	\thanks{
		Zhenzhou Jin, Li You, and Xiqi Gao are with the National Mobile Communications Research Laboratory, Southeast University, Nanjing 210096, China, and also with the Purple Mountain Laboratories, Nanjing 211100, China (e-mail: zzjin@seu.edu.cn; lyou@seu.edu.cn; xqgao@seu.edu.cn).
		
		Xiang-Gen Xia is with the Department of Electrical and Computer Engineering, University of Delaware, Newark, DE 19716, USA (e-mail: xxia@ee.udel.edu).
	}
%
%



			
	}

\maketitle

\begin{abstract}


The paradigm shift from environment-unaware communication to intelligent environment-aware communication is expected to facilitate the acquisition of channel state information for future wireless communications. Channel Fingerprint (CF), as an emerging enabling technology for environment-aware communication, provides channel-related knowledge for potential locations within the target communication area. However, due to the limited availability of practical devices for sensing environmental information and measuring channel-related knowledge, most of the acquired environmental information and CF are coarse-grained, insufficient to guide the design of wireless transmissions. To address this, this paper proposes a deep conditional generative learning approach, namely a customized \textit{conditional} generative diffusion model (CDiff). The proposed CDiff simultaneously refines environmental information and CF, reconstructing a fine-grained CF that incorporates environmental information, referred to as EnvCF, from its coarse-grained counterpart. Experimental results show that the proposed approach significantly improves the performance of EnvCF construction compared to the baselines.

\end{abstract}

\section{Introduction}\label{sec:net_intro}



Driven by the synergy of artificial intelligence (AI) and environmental sensing, 6G is poised to undergo a paradigm shift, evolving from environment-unaware communications to intelligent environment-aware ones \cite{zeng2024tutorial}. Channel fingerprint (CF) is an emerging enabling technology for environment-aware communications that provides location-specific channel knowledge for a potential base station (BS) in base-to-everything (B2X) pairs \cite{zeng2024tutorial,jin2024i2i}. Ideally, without considering the costs of sensing, computation, and storage, an ultra-fine-grained CF would encapsulate channel-related knowledge for all locations within the target communication area, such as channel gain and angle of arrival/departure, thereby alleviating the challenges of acquiring channel state information. By providing essential channel-related knowledge, CF has recently gained significant research attention for diverse applications, including object sensing, beamforming, localization \cite{zeng2024tutorial,jin2024i2i,yang2013rssi}.

A fundamental challenge for the aforementioned CF-based emerging applications is the construction of a sufficiently fine-grained CF, which is essential to ensure the accurate acquisition of channel-related information at specific locations. Existing related works can generally be categorized into model-based and data-driven approaches. In terms of model-based approaches, the authors of \cite{8662745} leverage prior assumptions about the physical propagation model along with partially measured channel data to estimate channel-related knowledge at potential locations. In \cite{10530520}, the authors utilize an analytical channel model to represent the spatial variability of the propagation environment, with channel modeling parameters estimated from measured data to reconstruct the CF. In data-driven approaches, one straightforward method for CF construction is interpolation-based, with classic techniques including radial basis function (RBF) \cite{zhang2024radiomap} interpolation and Kriging \cite{sato2017kriging}. Additionally, AI-based approaches for CF construction are recently emerging. In \cite{jin2024i2i}, the authors transform the CF estimation task into an image-to-image inpainting problem and develop a Laplacian pyramid-based model to facilitate CF construction. In \cite{9354041, 9771088}, UNet is utilized to learn geometry-based and physics-based features in urban or indoor environments, enabling the construction of corresponding CFs. In \cite{zeng2024tutorial}, fully connected networks are used to predict channel knowledge based on 2D coordinate locations. It is evident that CF construction is primarily influenced by the wireless propagation environment, but the nodes and devices available for sensing environmental information and measuring location-specific channel knowledge are usually limited in practice. Most existing methods focus on constructing CFs by either leveraging partial channel knowledge or incorporating prior assumptions about propagation models and wireless environment characteristics. However, limited work has been dedicated to simultaneously refining environmental information and channel-related knowledge, that is, reconstructing a finer-grained CF that integrates environmental information, referred to as environmental CF (EnvCF), from a coarse-grained one.

This paper investigates the construction of a fine-grained EnvCF in scenarios where environmental information and channel knowledge are limited by the high cost and availability constraints of sensing and testing equipment. Specifically, we reformulate the task of fine-grained EnvCF construction as an image super-resolution (ISR) problem. Considering that the conditional distribution of SR outputs given low-resolution (LR) inputs typically follows a complicated parametric distribution, leading to suboptimal performance of most feedforward neural network-based regression algorithms in ISR tasks \cite{Rombach_2022_CVPR}. To this end, leveraging the powerful implicit prior learning capability of generative diffusion model (GDM) \cite{Ho11,jin2024gdm4mmimo}, we propose a \textit{conditional} GDM (CDiff) to approximate the conditional distribution of the HR EnvCF. Specifically, within the variational inference framework, we derive an evidence lower bound (ELBO) of the log-conditional marginal distribution of the observed high-resolution (HR) EnvCF as the objective function. Furthermore, to make the transformation from a standard normal distribution to the target distribution more controllable, we incorporate the LR EnvCF as side information during the optimization of the denoising network. Simulation results show the superiority of the proposed approach.

\begin{figure*}[!b]
	\centering
	\includegraphics[width=0.96\textwidth]{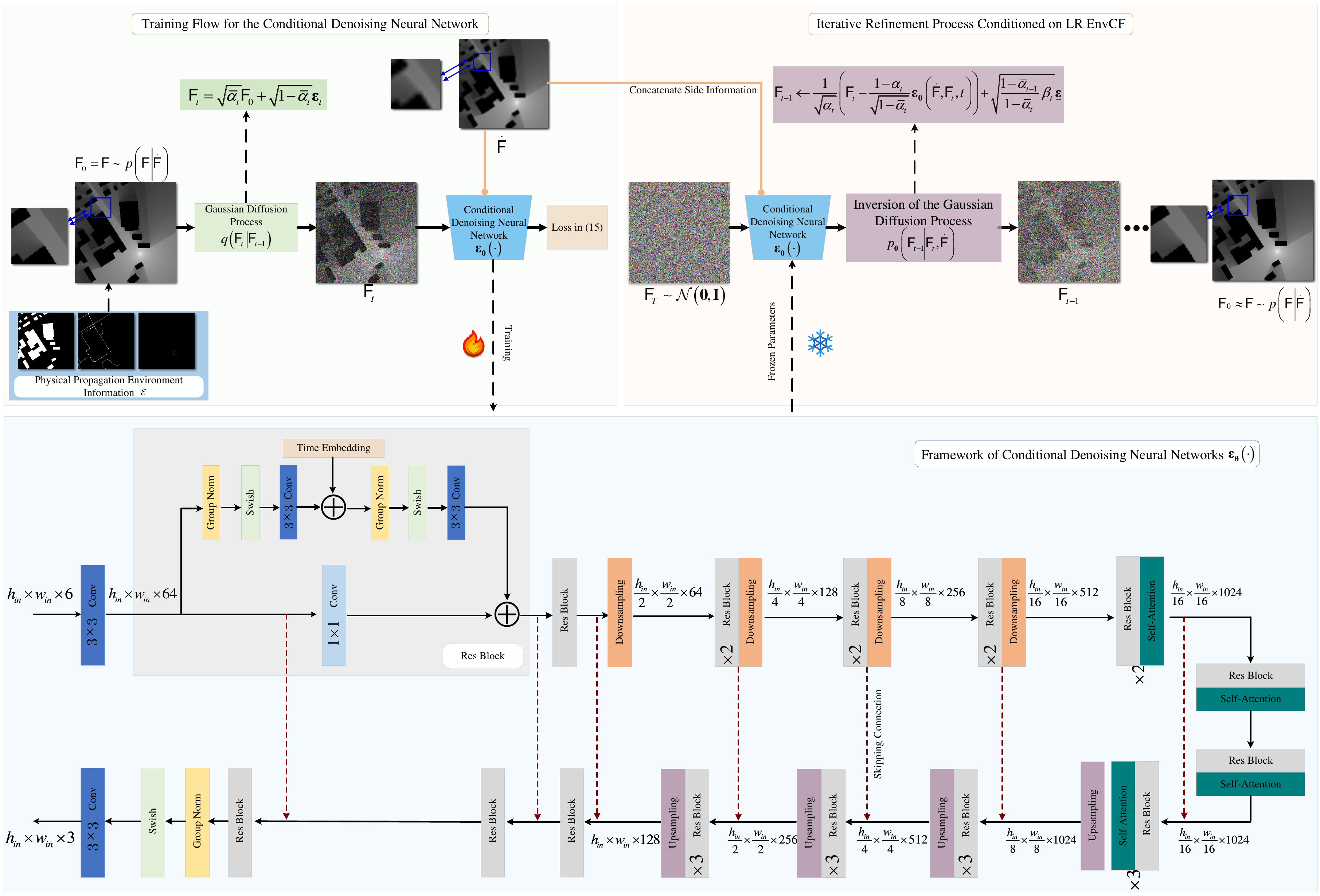}
	\captionsetup{font=footnotesize}
	\caption{Schematic of the proposed CDiff workflow and the architecture of the conditional denoising neural network.}
	\label{fig:model}
\end{figure*}

\section{EnvCF Model And Problem Formulation}\label{sec:sys_mod}
In this section, we first present the channel gain and the EnvCF model. We then reformulate the fine-grained EnvCF reconstruction problem by aligning the objective function with the learning of the gradient of the log-conditional density.


\subsection{EnvCF Model}
Consider a wireless communication scenario within a target area of interest, $A\subset {\mathbb{R}^2}$, where a base station (BS) serves $M$ user terminals (UTs), with their 2D position coordinates denoted as $\left\{ {{{{\mathbf{x}}_m}}} \right\}_{m = 1}^M = {\mathcal{A}}$. The attenuation of received signal power at UT is widely attributed to the physical characteristics of the wireless propagation environment, such as the geometric contours of real-world streets and buildings in urban maps, represented by $\mathcal{E}$. Key contributing factors include path loss along various propagation paths, reflections and diffractions caused by buildings, the ground, or other objects, waveguide effects in densely populated urban areas, and signal blockages from obstacles \cite{zeng2024tutorial}. The relatively slow-varying components of these influencing factors collectively form the channel gain function, denoted as ${\overline{g}}\left( \mathcal{E},\mathbf{x}_m\right)$, which reflects the large-scale signal attenuation measured at the UTs located at $\left\{ {{{{\mathbf{x}}_m}}} \right\}_{m = 1}^M = {\mathcal{A}}$. Additionally, small-scale effects are typically modeled as a complex Gaussian random variable ${\overline{{h}}}$ with unit variance. Without loss of generality, the baseband signal received by the UT at $ {{{{\mathbf{x}}_m}}}$ can be represented as \cite{jin2024i2i}
\begin{align}\label{eq:}
	\overline{{y}}\left(\mathcal{E},{{\mathbf{x}}_m} \right) = \sqrt {{\overline{g}}\left( \mathcal{E},\mathbf{x}_m\right)}{\overline{{h}}}{s}  + \overline{{{z}}}\left(\mathcal{E},{{\mathbf{x}}_m} \right),
\end{align}
where ${s}$ represents the transmitted signal with power ${P_{\cal X}}$, and $\overline{{{z}}}\left(\mathcal{E},{{\mathbf{x}}_m} \right)$ denotes the additive noise with a single-sided power spectral density of ${N_0}$. The average received energy per symbol can be expressed as 
\begin{align}\label{eq:}
	{{P_{\cal Y}}}=\frac{{\mathbb{E}[{{\left| \overline{{y}}\left(\mathcal{E},{{\mathbf{x}}_m} \right) \right|}^2}]}}{B} = \frac{{{\overline{g}}\left( \mathcal{E},\mathbf{x}_m\right){{P_{\cal X}}}}}{B} + {N_0},
\end{align}
where $B$ denotes the signal bandwidth. The channel gain, in dB, for the UT located at ${{\mathbf{x}}_m}$ is defined as
\begin{align}\label{eq:2}
	{G}\left(\mathcal{E},\mathbf{x}_m\right) = {\left( {{P_{\cal Y}}}\right) _{\rm{dB}}} - {\left( {P_{\cal X}}\right) _{\rm{dB}}}.
\end{align}
The channel gain ${G}\left(\mathcal{E},\mathbf{x}_m\right)$ in \eqref{eq:2} is primarily influenced by the propagation environment and the location of the UT. 


Assuming the target area of interest has a size of $\mathcal{D} \times \mathcal{D}$, we perform spatial discretization along both the $X$-axis and the $Y$-axis. Specifically, a resolution factor $\delta$ is defined, with the minimum spacing units for the spatial discretization process set as $\Delta_x={\mathcal{D} \mathord{\left/
		{\vphantom {\mathcal{D} \delta}} \right.
		\kern-\nulldelimiterspace} \delta}$ and $\Delta_y={\mathcal{D} \mathord{\left/
		{\vphantom {\mathcal{D} \delta}} \right.
		\kern-\nulldelimiterspace} \delta}$. Each spatial grid is denoted as ${{\mathbf{\Upsilon }}_{i,j}}$, where $i = 1,2,...,	{\mathcal{D} \mathord{\left/
		{\vphantom {\mathcal{D} \Delta_x}} \right.
		\kern-\nulldelimiterspace} \Delta_x}$ and $j = 1,2,...,	{\mathcal{D} \mathord{\left/
		{\vphantom {\mathcal{D} \Delta_y}} \right.
		\kern-\nulldelimiterspace} \Delta_y}$, and the $(i,j)$-th spatial grid is given by
\begin{align}\label{eq:}
	{{\mathbf{\Upsilon }} _{i,j}}: = {[i{\Delta _x},j{\Delta {}_y]^T}}.
\end{align}
Through the spatial discretization process, the physical propagation environment information $\mathcal{E}$ of the target area $A$ can be rearranged into a two-dimensional tensor, defined as ${\boldsymbol{\mathsf{E}}}$, i.e., ${\left[ {{\boldsymbol{\mathsf{E}}}} \right]_{i,j}} = \mathcal{E}\left( {{{\mathbf{\Upsilon }}_{i,j}}} \right)$. Similarly, the channel gains collected at potential UT locations within $A$ are rearranged into a tensor with an image-like structure, referred to as CF, defined as ${\left[ {\boldsymbol{\mathsf{G}}} \right]_{i,j}} = G\left( {{\left[ {{\boldsymbol{\mathsf{E}}}} \right]_{i,j}},{{\mathbf{\Upsilon }}_{i,j}}} \right)$. Then, EnvCF, i.e., CF, is integrated with the wireless propagation environment, defined as
\begin{align}
	{[{\boldsymbol{\mathsf{F}}}]_{i,j}} = G\left( {{\left[ {{\boldsymbol{\mathsf{E}}}} \right]_{i,j}},{{\mathbf{\Upsilon }}_{i,j}}} \right)+{\left[ {{\boldsymbol{\mathsf{E}}}} \right]_{i,j}},
\end{align}
where ${\boldsymbol{\mathsf{E}}}$ represents the global propagation environment. Note that in our simulation process, ${\boldsymbol{\mathsf{E}}}$ represents an urban map with BS location, stored as a morphological 2D image. Binary pixel values of 0 and 1 are utilized to depict buildings and streets with various shapes and geometric layouts, as well as the location of BS. $\mathcal{E}\left({{\mathbf{\Upsilon }}_{i,j}}\right)$ represents the local propagation environment at each square meter (or each pixel). The channel gain $G\left( {{\left[ {{\boldsymbol{\mathsf{E}}}} \right]_{i,j}},{{\mathbf{\Upsilon }}_{i,j}}} \right)$ is computed using the professional channel simulation software \textit{WinProp} \cite{9354041,jin2024i2i} and then converted into grayscale pixel values ranging from 0 to 1. Therefore, EnvCF is modeled as a 2D environmental channel gain map, comprising both the propagation environment map and the channel gains at each UT location, as illustrated by one of the EnvCF samples presented in \figref{fig:model}. It can be observed that when factors such as time and frequency are considered, the EnvCF model can be extended to a multi-dimensional tensor.



\subsection{Problem Formulation}

It is evident that a finer-grained EnvCF can provide more accurate information about the physical environment and channel gains, which is beneficial for wireless transmission design \cite{jin}. However, the EnvCF obtained by practical BS is typically coarse-grained due to the limited availability for collecting environmental information and channel knowledge at specific locations. Therefore, our task focuses on refining both environmental and channel gain information from a given coarse-grained EnvCF, particularly in scenarios constrained by sensing costs, implicit limitations, or security concerns.

Define a low-resolution (LR) factor $\delta_{\mathrm{LR}}$ and a high-resolution (HR) factor $\delta_{\mathrm{HR}}$. Correspondingly, the LR EnvCF and HR EnvCF are represented as ${\boldsymbol{\mathsf{F}}}_{\mathrm{LR}}$ and ${\boldsymbol{\mathsf{F}}}_{\mathrm{HR}}$, respectively, which are collected and rearranged by discretizing the target area into $\delta_{\mathrm{LR}}\times\delta_{\mathrm{LR}}$ and $\delta_{\mathrm{HR}}\times\delta_{\mathrm{HR}}$ grids, respectively. Then, our task is to establish a mapping capable of reconstructing a HR EnvCF from a given LR EnvCF, expressed as 
\begin{align}\label{eq:6}
	{\mathcal{M}_\Theta }:\boldsymbol{\mathsf{F}}_{{\rm{LR}},n} \to \boldsymbol{\mathsf{F}}_{{\rm{HR}},n},\quad \forall n \in \left\{ {1,2,\ldots,N} \right\},
\end{align}
where $\Theta$ denotes the learnable parameters for the mapping $\mathcal{M}_\Theta$, while $N$ indicates the number of training samples. However, \eqref{eq:6} is an undetermined inverse problem. Given that the conditional distribution of HR outputs for a given LR input rarely adheres to a simple parametric distribution, most regression methods based on feedforward neural networks for \eqref{eq:6} tend to struggle with high upscaling factors, often failing to reconstruct fine details accurately \cite{Rombach_2022_CVPR}. Fortunately, GDM has proven effective in capturing the complex empirical distributions of target data. Specifically, if the implicit prior information of the HR EnvCF distribution, such as the gradient of the data log-density, can be effectively learned, it becomes possible to transition to the target EnvCF distribution through iterative refinement steps from a standard normal distribution, akin to Langevin dynamics. Meanwhile, the ``noise'' estimated in traditional GDM is equivalent to the gradient of the data log-density. Therefore, \eqref{eq:6} can be further formulated as 
\begin{subequations}\label{eq:7}
	\begin{align}
		&\!\!\!\mathop {{\text{argmin}}}\limits_\Theta  {{\mathbb{E}}_{{P}(\boldsymbol{\mathsf{F}}_\mathrm{H\!R},\boldsymbol{\mathsf{F}}_\mathrm{L\!R})\!\!}}\!\left[ {{{\!\left\| {\nabla \!\log\! {P}(\!\boldsymbol{\mathsf{F}}_\mathrm{H\!R}|\boldsymbol{\mathsf{F}}_\mathrm{L\!R}\!) \!\!-\!\!\! \nabla \!\log\! {P_{\Theta}}(\!\boldsymbol{\mathsf{F}}_\mathrm{H\!R}|\boldsymbol{\mathsf{F}}_\mathrm{L\!R}\!)} \right\|}^2_2}} \right]\!\!,\!\label{eq:dbso}\\
		&\!\!\!\qquad\qquad\quad{\rm{s.t.}}\text{  } {\mathbf{x}_m} \in {\mathcal{A}} ,n \in \{1,2,\ldots, N\},\label{eq:}
	\end{align}
\end{subequations}
where $\nabla \log {P}(\cdot)$ represents the gradient of the log-density, and $P_{\Theta}$ denotes the learned density. To this end, leveraging the powerful implicit prior learning capability of GDM, we tailor the CDiff to solve \eqref{eq:7}, with the detailed implementation provided in Sec. \ref{sec:CGDM}. For simplicity, ${\boldsymbol{\mathsf{F}}}_{\mathrm{LR}}$ and ${\boldsymbol{\mathsf{F}}}_{\mathrm{HR}}$ are represented by ${ \boldsymbol{\dot{\mathsf{F}}}}$ and $\boldsymbol{\mathsf{F}}$, respectively, in the subsequent sections.

\section{HR EnvCF Reconstruction via CDiff}\label{sec:CGDM}

Depending on the resolution factors $\delta_{\mathrm{LR}}$ and $\delta_{\mathrm{HR}}$, the sensing nodes and measurement devices deployed in practice to collect channel knowledge and environmental information can acquire the corresponding LR EnvCF samples, ${\boldsymbol{\dot{\mathsf{F}}}}_n$, and HR EnvCF samples, $\boldsymbol{\mathsf{F}}_n$. These samples form a paired LR-HR EnvCF dataset for training, denoted as $\mathcal{S}\!=\!\left\lbrace {{ \boldsymbol{\dot{\mathsf{F}}}}_n,\boldsymbol{\mathsf{F}}_n}\right\rbrace_{n=1}^N $, which is generally sampled from an unknown distribution $p\left({ \boldsymbol{\dot{\mathsf{F}}}},\boldsymbol{\mathsf{F}} \right)$. In our task, the goal is to learn a parametric approximation of $p\left( {\boldsymbol{\mathsf{F}}}\left| { \boldsymbol{\dot{\mathsf{F}}}} \right.\right)$ through a directed iterative refinement process, guided by side information in the form of LR EnvCF.

\subsection{Initiation of Gaussian Diffusion Process with HR EnvCF}

Denoted as ${\boldsymbol{\mathsf{F}}_0}={\boldsymbol{\mathsf{F}}} \sim q\left( \boldsymbol{\mathsf{F}} \right)$, the GDM employs a fixed diffusion process $q\left( {{\boldsymbol{\mathsf{F}}_{1:T}}\left| {{\boldsymbol{\mathsf{F}}_0}} \right.} \right)$, represented as a deterministic Markovian chain where Gaussian noise is gradually introduced to the sample over $T$ steps \cite{Ho11}:
\begin{subequations}
 \begin{align}
	q\left( {{\boldsymbol{\mathsf{F}}_{1:T}}\left| {{\boldsymbol{\mathsf{F}}_0}} \right.} \right) &= \prod\limits_{t = 1}^T {q\left( {{\boldsymbol{\mathsf{F}}_t}\left| {{\boldsymbol{\mathsf{F}}_{t - 1}}} \right.} \right)},\label{eq:chmd}\\
	q\left( {{\boldsymbol{\mathsf{F}}_t}\left| {{\boldsymbol{\mathsf{F}}_{t - 1}}} \right.} \right) &= \mathcal{N}\left( {{\boldsymbol{\mathsf{F}}_t};\sqrt {1 - {\beta _t}} {\boldsymbol{\mathsf{F}}_{t - 1}},{\beta _t}\mathbf{I}} \right),\label{eq:marginal ddpm}  \\
	{\boldsymbol{\mathsf{F}}_t} 
	&= \sqrt {1 - {\beta _t}} {\boldsymbol{\mathsf{F}}_{t - 1}} + \sqrt {{\beta _t}} {\boldsymbol{\varepsilon}},\label{eq:Gt}
 \end{align}
\end{subequations}
where $\left\{ {{\beta _t} \in \left( {0,1} \right)} \right\}_{t = 1}^T$ is a variance schedule that controls the noise level at each step, and $\boldsymbol{\varepsilon}$ denotes Gaussian noise following the distribution $\mathcal{N}\left( {\boldsymbol{\varepsilon}; {\mathbf{0},\mathbf{I}} } \right)$. Utilizing the reparameterization trick, $\boldsymbol{\mathsf{F}}_t$ can be sampled in a closed form as:
\begin{subequations}
\begin{align}
	\label{eq:9a}
	{{\boldsymbol{\mathsf{F}}}_t} 
	&=\sqrt {{{\overline \alpha  }_t}} {{\boldsymbol{\mathsf{F}}}_0} + \sqrt {1 - {{\overline \alpha  }_t}} {\boldsymbol{\varepsilon}}_t, \\
	\label{eq:chmd}
	q\left( {{\boldsymbol{\mathsf{F}}_t}\left| {{\boldsymbol{\mathsf{F}}_0}} \right.} \right) &= \mathcal{N}\left( {{\boldsymbol{\mathsf{F}}_t};\sqrt {{{\overline \alpha  }_t}} {\boldsymbol{\mathsf{F}}_0},\left(1 - {{\overline \alpha  }_t}\right)  \mathbf{I}} \right),
\end{align}
\end{subequations}
where ${\boldsymbol{\varepsilon}}_t \sim \mathcal{N}\left( { {\mathbf{0},\mathbf{I}} } \right)$, ${\alpha _t} = 1 - {\beta _t}$ and ${\overline \alpha  _t} = \prod\nolimits_{i = 1}^t {{\alpha_i}}$. Typically, the variance schedule is set as ${\beta _1} < {\beta _2} < ... < {\beta _T}$ \cite{JINTCCN}. As ${\beta _T}$ approaches 1, ${{\boldsymbol{\mathsf{F}}}_T}$ approximates a standard Gaussian distribution regardless of the initial state ${{{\boldsymbol{\mathsf{F}}}_0}}$, i.e., $q\left( {{{\boldsymbol{\mathsf{F}}}_T}\left| {{{\boldsymbol{\mathsf{F}}}_0}} \right.} \right) \approx \mathcal{N}\left( {{{\boldsymbol{\mathsf{F}}}_T};\mathbf{0},{\mathbf{I}}} \right)$.

\subsection{Inversion of Diffusion Process Conditioned on LR EnvCF}
In the proposed CDiff, the inversion process is viewed as a conditional decoding procedure, where at each time step $t$, ${\boldsymbol{\mathsf{F}}_{t}}$, conditioned on ${ \boldsymbol{\dot{\mathsf{F}}}}$, is denoised and refined to ${\boldsymbol{\mathsf{F}}_{t-1}}$, with the conditional transition probability for each step denoted as ${p\left( {{\boldsymbol{\mathsf{F}}_{t-1}}\left| {{\boldsymbol{\mathsf{F}}_{t}}},{ \boldsymbol{\dot{\mathsf{F}}}} \right.} \right)}$. Then, the conditional joint distribution of the inversion process is expressed as
\begin{align}
	\label{eq:10}
	{p}\left( {{{\boldsymbol{\mathsf{F}}}_{0:T}}} \left| {{ \boldsymbol{\dot{\mathsf{F}}}}} \right. \right) = {p}\left( {{{\boldsymbol{\mathsf{F}}}_T}} \right)\prod\limits_{t = 1}^T {{p}\left( {{{\boldsymbol{\mathsf{F}}}_{t - 1}}\left| {{{\boldsymbol{\mathsf{F}}}_t}, { \boldsymbol{\dot{\mathsf{F}}}}} \right.} \right)}.
\end{align}
To execute the conditional inversion process \eqref{eq:10}, a denoising neural network ${{{{{\boldsymbol{\varepsilon}}_{\boldsymbol{\theta }}} }}}\left( \cdot \right)$ with a learnable parameter set ${{\boldsymbol{\theta}}}$ needs to be designed to approximate the conditional transition densities:
\begin{subequations}
\begin{align}
	\label{eq:}
	{p}\left( {{{\boldsymbol{\mathsf{F}}}_{0:T}}} \left| {{ \boldsymbol{\dot{\mathsf{F}}}}} \right. \right) &= {p}\left( {{{\boldsymbol{\mathsf{F}}}_T}} \right)\prod\limits_{t = 1}^T {{p_{\boldsymbol{\theta }}}\left( {{{\boldsymbol{\mathsf{F}}}_{t - 1}}\left| {{{\boldsymbol{\mathsf{F}}}_t}, { \boldsymbol{\dot{\mathsf{F}}}}} \right.} \right)},\\
	\label{eq:}
	{p_{{\boldsymbol{\theta}}} }\!\left(\! {{{\boldsymbol{\mathsf{F}}}_{t \!-\! 1}}\!\left| {{{\boldsymbol{\mathsf{F}}}_t},\!{ \boldsymbol{\dot{\mathsf{F}}}}} \right.} \!\right) &= \mathcal{N}\!\left(\! {{{\boldsymbol{\mathsf{G}}}_{t \!-\! 1}};\!{{\boldsymbol{\mu}}_{{\boldsymbol{\theta}}} }\left({ \boldsymbol{\dot{\mathsf{F}}}}, {{{\boldsymbol{\mathsf{F}}}_t},t} \right)\!,\!\mathbf{\Sigma}_{{\boldsymbol{\theta}}} {\left({ \boldsymbol{\dot{\mathsf{F}}}},\! {{{\boldsymbol{\mathsf{F}}}_t},\!t} \right)} } \!\right)\!\!.
\end{align}
\end{subequations}
Note that the proposed CDiff incorporates a denoising neural network ${{{{{\boldsymbol{\varepsilon}}_{\boldsymbol{\theta }}} }}}\left( \cdot \right)$ conditioned on side information in the form of an LR EnvCF ${\boldsymbol{\dot{\mathsf{F}}}}$, guiding it to progressively denoise from a Gaussian-distributed ${\boldsymbol{\mathsf{F}}_T}$ and generate the HR EnvCF ${\boldsymbol{\mathsf{F}}_0}$. 
\begin{algorithm}[!t]
	\caption{Training Strategy for the Conditional Denoising Neural Network}
	\label{alg:train}
	\begin{algorithmic}[1]
		\REPEAT
		\STATE
		Load data pairs $\left({ \boldsymbol{\dot{\mathsf{F}}}},\boldsymbol{\mathsf{F}}_0 \right) \sim p\left({ \boldsymbol{\dot{\mathsf{F}}}},\boldsymbol{\mathsf{F}}_0 \right) $ from the EnvCF training dataset $\mathcal{S}\!=\!\left\lbrace {{ \boldsymbol{\dot{\mathsf{F}}}}_n,\boldsymbol{\mathsf{F}}_n}\right\rbrace_{n=1}^N $ 
		\STATE
		Sample time steps $t$ uniformly from ${1, \dots, T}$, i.e., $t \sim {\rm{Uniform}}({1,...,T})$
		\STATE
		Generate a noise tensor ${\boldsymbol{\varepsilon}_t}$ of the same dimensions as $\boldsymbol{\mathsf{F}}_0$, ${\boldsymbol{\varepsilon}_t} \sim \mathcal{N}\left( {\mathbf{0},\mathbf{I}} \right)$
		\STATE
		Perform the diffusion process on the HR EnvCF $\boldsymbol{\mathsf{F}}_0$, ${{\boldsymbol{\mathsf{F}}}_t} =\sqrt {{{\overline \alpha  }_t}} {{\boldsymbol{\mathsf{F}}}_0} + \sqrt {1 - {{\overline \alpha  }_t}} {\boldsymbol{\varepsilon}}_t$
		\STATE Feed ${{{\boldsymbol{\mathsf{F}}}_t}}$, ${\boldsymbol{\dot{\mathsf{F}}}}$, and $t$ into the network ${{{{{\boldsymbol{\varepsilon}}_{\boldsymbol{\theta }}} }}}\left( \cdot \right)$\STATE
		Perform gradient descent step on the loss function \eqref{eq:15} to optimize the model parameters ${\boldsymbol{\theta }}$:\\
		\quad $ \nabla_{\boldsymbol{\theta }} \big\| {{\boldsymbol{\varepsilon }}_t} - {{\boldsymbol{\varepsilon }}_{\boldsymbol{\theta }}}\big({{ \boldsymbol{\dot{\mathsf{F}}}}}, {\sqrt {{{\bar \alpha }_t}} {{{\boldsymbol{\mathsf{F}}}_0}} \!+\! \sqrt {1 - {{\bar \alpha }_t}} {{\boldsymbol{\varepsilon }}}},t \big) \big\|_2^2 $
		\UNTIL the loss function \eqref{eq:15} converges
	\end{algorithmic}
\end{algorithm}
\begin{algorithm}[!t]
	\caption{HR EnvCF Generation via $T$-Step Conditional Inversion Diffusion Process}
	\label{alg:infer}
	\begin{algorithmic}[1]
		\STATE
		Load the trained model and its weight set ${\boldsymbol{\theta }}$
		\STATE
		Obtain ${{{\boldsymbol{\mathsf{F}}}_T}}\sim \mathcal{N}\left( {\mathbf{0},\mathbf{I}} \right)$, and ${\boldsymbol{\dot{\mathsf{F}}}}$
		\FOR{$t=T,..,1$}
		\STATE
		${\underline{\boldsymbol{\varepsilon}}} \sim\mathcal{N}\left( {\mathbf{0},\mathbf{I}} \right)$ if $t>1$, else ${\underline{\boldsymbol{\varepsilon}}} =0$
		\STATE
		Execute the conditional iterative refinement step:\\
		\quad ${{{\boldsymbol{\mathsf{F}}}_{t\!-\!1}}} \!\leftarrow\! \frac{1}{{\sqrt {{\alpha _t}} }}\!\left(\! {{{{\boldsymbol{\mathsf{F}}}}_t} \!-\! \frac{{1 \!-\! {\alpha _t}}}{{\sqrt {1 \!-\! {{\bar \alpha }_t}} }}{{\boldsymbol{\varepsilon}}_{\boldsymbol{\theta }}}\!\left(\!{{ \boldsymbol{\dot{\mathsf{F}}}}, {{\boldsymbol{\mathsf{F}}}_t},t} \!\right)} \!\right) \!+\! \sqrt { \frac{{1 \!-\! {{\bar \alpha }_{t \!-\! 1}}}}{{1 \!-\! {{\bar \alpha }_t}}}{\beta _t}} {\underline{\boldsymbol{\varepsilon}}}$
		\ENDFOR
		\RETURN ${{\hat{{\boldsymbol{\mathsf{F}}}}_0}}$
		
	\end{algorithmic}
\end{algorithm}

To ensure the effective functioning of this denoising neural network ${{{{{\boldsymbol{\varepsilon}}_{\boldsymbol{\theta }}} }}}\left( \cdot \right)$, a specific objective function needs to be derived. One common likelihood-based approach in generative modeling involves optimizing the model to maximize the conditional joint probability distribution $p\left( {{{{\boldsymbol{\mathsf{F}}}}_{0:T}}}\left| {{ \boldsymbol{\dot{\mathsf{F}}}}} \right. \right)$ of all observed samples. However, only the observed sample ${\boldsymbol{\mathsf{F}}_0}$ is accessible, while the latent variables ${{{\boldsymbol{\mathsf{F}}}_{1:T}}}$ remain unknown. To this end, we seek to maximize the conditional marginal distribution $p\left( {{{{\boldsymbol{\mathsf{F}}}}_{0}}}\left| {{ \boldsymbol{\dot{\mathsf{F}}}}} \right. \right)$, expressed as
\begin{align}
	p\left( {{{\boldsymbol{\mathsf{F}}}_0}}\left| {{ \boldsymbol{\dot{\mathsf{F}}}}} \right. \right) = \int {p\left( {{{\boldsymbol{\mathsf{F}}}_{0:T}}}\left| {{ \boldsymbol{\dot{\mathsf{F}}}}} \right. \right)} d{{\boldsymbol{\mathsf{F}}}_{1:T}}.
\end{align}
By leveraging variational inference techniques, we can derive the evidence lower bound (ELBO) as a surrogate objective to optimize the denoising neural network:
\begin{align}
	    \label{eq:13b}
		\log p\left( {{{\boldsymbol{\mathsf{F}}}_0}}\left| {{ \boldsymbol{\dot{\mathsf{F}}}}} \right. \right) 
		\mathop  \geqslant \limits^{(\mathrm{a})} {\mathbb{E}_{q\left( {{{\boldsymbol{\mathsf{F}}}_{1:T}}\left| {{{\boldsymbol{\mathsf{F}}}_0}} \right.} \right)}}\left( {\log \frac{{p\left( {{{\boldsymbol{\mathsf{F}}}_{0:T}}}\left| {{ \boldsymbol{\dot{\mathsf{F}}}}} \right. \right)}}{{q\left( {{{\boldsymbol{\mathsf{F}}}_{1:T}}\left| {{{\boldsymbol{\mathsf{F}}}_0}} \right.} \right)}}} \right), 
\end{align}
where $\mathop{\geqslant}\limits^{(\mathrm{a})}$ in \eqref{eq:13b} follows from Jensen's inequality. Then, \eqref{eq:13b} can be further expressed as \eqref{eq:14}, displayed at the top of the next page.
\begin{figure*}[ht] 
	\begin{subequations}\label{eq:14}
		\begin{align}\label{eq:35}
			&\!\!\!\!\!\log p\left( {{{\boldsymbol{\mathsf{F}}}_0}}\left| {{ \boldsymbol{\dot{\mathsf{F}}}}} \right. \right)
			\geqslant {\mathbb{E}_{q\left( {{{\boldsymbol{\mathsf{F}}}_{1:T}}\left| {{{\boldsymbol{\mathsf{F}}}_0}} \right.} \right)}}\left( {\log \frac{{p\left( {{{\boldsymbol{\mathsf{F}}}_T}} \right){p_{{\boldsymbol{\theta}}}}\left( {{{\boldsymbol{\mathsf{F}}}_0}\left| {{{\boldsymbol{\mathsf{F}}}_1},{ \boldsymbol{\dot{\mathsf{F}}}} } \right.} \right)}}{{q\left( {{{\boldsymbol{\mathsf{F}}}_1}\left| {{{\boldsymbol{\mathsf{F}}}_0}} \right.} \right)}} + \log \frac{{q\left( {{{\boldsymbol{\mathsf{F}}}_1}\left| {{\boldsymbol{\mathsf{F}}_0}} \right.} \right)}}{{q\left( {{\boldsymbol{\mathsf{F}}_T}\left| {{{\boldsymbol{\mathsf{F}}}_0}} \right.} \right)}} + \log \prod\limits_{t = 2}^T {\frac{{{p_{{\boldsymbol{\theta}}}}\left( {{{\boldsymbol{\mathsf{F}}}_{t - 1}}\left| {{{\boldsymbol{\mathsf{F}}}_t},{ \boldsymbol{\dot{\mathsf{F}}}} } \right.} \right)}}{{q\left( {{{\boldsymbol{\mathsf{F}}}_{t - 1}}\left| {{{\boldsymbol{\mathsf{F}}}_t},{{\boldsymbol{\mathsf{F}}}_0}} \right.} \right)}}} } \right)\\
			&\!\!\!\!\!=  {{\mathbb{E}_{q\left( {{{\boldsymbol{\mathsf{F}}}_1}\left| {{{\boldsymbol{\mathsf{F}}}_0}} \right.} \right)}}\!\left(\! {\log {p_{{\boldsymbol{\theta}}}}\!\left(\! {{{\boldsymbol{\mathsf{F}}}_0}\!\left| {{{\boldsymbol{\mathsf{F}}}_1},\!{ \boldsymbol{\dot{\mathsf{F}}}} } \right.} \!\right)} \!\right)}\!\!-\!\!  {\sum\limits_{t = 2}^T {{\mathbb{E}_{q\left( {{{\boldsymbol{\mathsf{F}}}_t}\left| {{{\boldsymbol{\mathsf{F}}}_0}} \right.} \right)}}\!\left(\! {{D_{{\text{KL}}}}\!\left(\! {\left. {q\!\left( {{{\boldsymbol{\mathsf{F}}}_{t \!-\! 1}}\!\left| {{{\boldsymbol{\mathsf{F}}}_t},\!{{\boldsymbol{\mathsf{F}}}_0}} \right.} \!\right)} \right\|{p_{\boldsymbol{\theta }}}\!\left(\! {{{\boldsymbol{\mathsf{F}}}_{t\! - \!1}}\!\left| {{{\boldsymbol{\mathsf{F}}}_t},\!{ \boldsymbol{\dot{\mathsf{F}}}} } \right.} \!\!\right)} \!\!\right)} \!\!\right)} } \!\!-\!\!  {{D_{\text{KL}}}\!\left(\! {\left. {q\!\left(\! {{{\boldsymbol{\mathsf{F}}}_T}\!\left| {{{\boldsymbol{\mathsf{F}}}_0}} \right.}\! \right)} \right\|\!p\!\left(\! {{{\boldsymbol{\mathsf{F}}}_T}} \!\right)} \!\right)} \label{eq:35e}
		\end{align}
	\end{subequations}
	\hrule
\end{figure*}
Here, ${D_{\text{KL}}}( \cdot )$ represents the Kullback-Leibler (KL) divergence. The components of the ELBO \eqref{eq:14} for the log-conditional marginal distribution are similar to those in the surrogate objective presented in \cite{Ho11}. By applying Bayes' theorem, the final simplified objective can be derived as
\begin{align}\label{eq:15} 
	\!\!\!\!\! { {\mathcal{L}}} (\boldsymbol{\theta}) \!:=\!\! \sum\limits_{t = 1}^T\!{\mathbb{E}_{{{{{\boldsymbol{\mathsf{F}}}_0}}}\!,{{\boldsymbol{\varepsilon }}_t} }}\!\Big( \!\big\| {{\boldsymbol{\varepsilon }}_t} \!\!-\!\! {{\boldsymbol{\varepsilon }}_{\boldsymbol{\theta }}}\big({{ \boldsymbol{\dot{\mathsf{F}}}}},\! \underbrace{\sqrt {{{\bar \alpha }_t}} {{{\boldsymbol{\mathsf{F}}}_0}} \!+\! \sqrt {1 \!-\! {{\bar \alpha }_t}} {{\boldsymbol{\varepsilon }}}}_{{{\boldsymbol{\mathsf{F}}}_t}},\!t \big) \big\|_2^2 \Big)\!.\!\!\!\!\!\!\!
\end{align}
Based on the trained CDiff, given the noise-contaminated EnvCF ${{{\boldsymbol{\mathsf{F}}}_t}}$ and the side information ${{ \boldsymbol{\dot{\mathsf{F}}}}}$, we can approximate the target HR EnvCF $\boldsymbol{\mathsf{F}}_0$ through \eqref{eq:9a}, i.e., 
\begin{align}\label{eq:}
	{{\hat{{\boldsymbol{\mathsf{F}}}}_0}} \!=\! \frac{1}{{\sqrt {{{\bar \alpha }_t}} }}\!\!\left(\!\! {{{{\boldsymbol{\mathsf{F}}}_t}} \!-\! \sqrt {1 \!-\! {{\bar \alpha }_t}} {{\boldsymbol{\varepsilon }}_{\boldsymbol{\theta }}}\big({{ \boldsymbol{\dot{\mathsf{F}}}}}, \underbrace{\sqrt {{{\bar \alpha }_t}} {{{\boldsymbol{\mathsf{F}}}_0}} \!+\! \sqrt {1 \!-\! {{\bar \alpha }_t}} {{\boldsymbol{\varepsilon }}}}_{{{\boldsymbol{\mathsf{F}}}_t}},t \big)} \!\!\right)\!.
\end{align}
Each iteration in the proposed CDiff is expressed as
\begin{align}\label{eq:}
	\!\!\!\!\!{{{\boldsymbol{\mathsf{F}}}_{t\!-\!1}}} \!\leftarrow\! \frac{1}{{\sqrt {{\alpha _t}} }}\!\!\left(\!\! {{{{\boldsymbol{\mathsf{F}}}}_t} \!\!-\!\! \frac{{1 \!-\! {\alpha _t}}}{{\sqrt {1 \!-\! {{\bar \alpha }_t}} }}{{\boldsymbol{\varepsilon}}_{\boldsymbol{\theta }}}\!\left(\!{{ \boldsymbol{\dot{\mathsf{F}}}}, {{\boldsymbol{\mathsf{F}}}_t},t} \!\right)} \!\!\right) \!+\! \sqrt { \frac{{1 \!-\! {{\bar \alpha }_{t \!-\! 1}}}}{{1 \!-\! {{\bar \alpha }_t}}}{\beta _t}} {\underline{\boldsymbol{\varepsilon}}},\!\!\!\!\!\!
\end{align}
where ${\underline{\boldsymbol{\varepsilon}}} \sim\mathcal{N}\left( {\mathbf{0},\mathbf{I}} \right)$. For clarity, the training process and the iterative inference process of the proposed CDiff are summarized in \textbf{\alref{alg:train}} and \textbf{\alref{alg:infer}}, respectively.

\section{Numerical Experiment}

In this section, we first present the generation of the EnvCF dataset and the parameter configuration of the proposed model. Then, we conduct both quantitative and qualitative comparisons with the baselines on the $\times4$ EnvCF reconstruction task (i.e., $64 \times 64 \to 256 \times 256$).

\newcolumntype{L}{>{\hspace*{-\tabcolsep}}l}
\newcolumntype{R}{c<{\hspace*{-\tabcolsep}}}
\definecolor{lightblue}{rgb}{0.93,0.95,1.0}
\begin{table}[!b]
	\captionsetup{font=footnotesize}
	\caption{System and Model Parameters}\label{ta:sys}
	\centering
	\setlength{\tabcolsep}{13mm}
	\ra{2}
	\scriptsize
	\scalebox{0.8}{\begin{tabular}{LR}
			\toprule
			Parameter &  Value\\
			\midrule
			\rowcolor{lightblue}
			Size of the target area $A$ & $\mathcal{D} \times \mathcal{D}=256\times256$ $\mathrm{m}^2$ \\
			Sampling interval for HR EnvCF & $\Delta_x=\Delta_y=1$ $\mathrm{m}$\\
			\rowcolor{lightblue}
			Sampling interval for LR EnvCF & $\Delta_x=\Delta_y=4$ $\mathrm{m}$\\
			Carrier frequency & $f=5.9$ GHz\\
			\rowcolor{lightblue}
			Bandwidth & $B=10$ MHz\\
			Transmit power & 23 dBm\\
			\rowcolor{lightblue}
			Noise power sepctral density & -174 dBm/Hz\\
			Variance schedule  & Linear: $\beta _1=10^{-6}  \to \beta _T= 10^{-2}$\\
			\bottomrule
		\end{tabular}
	}
\end{table}

\subsection{Datasets and Experiment Setup }
The RadioMapSeer dataset \cite{9354041}, a widely adopted CF dataset that incorporates environmental information, is utilized for training and validating the proposed CDiff. Specifically, the RadioMapSeer dataset consists of 700 unique city maps, each measuring $256 \times 256$ $\rm{m}^2$ and containing 80 distinct BS locations. For each possible combination of city maps and BS locations, the dataset provides the corresponding CFs, which are simulated using \textit{WinProp} \cite{9354041,jin2024i2i}. These city maps describe the geometric contours of real-world streets and buildings, sourced from OpenStreetMap \cite{9354041} for different cities. Considering a highly challenging $\times4$ EnvCF refinement task, we set $\delta_{\mathrm{HR}}=256$, utilizing a sampling interval of $\Delta_x=\Delta_y=1$ m to sample the city map along with its associated channel gain values and environmental information, resulting in the HR EnvCF, denoted as $\boldsymbol{\mathsf{F}}_{{\rm{HR}},n}$. For the LR counterpart, $\delta_{\mathrm{LR}}$ is set to 64, meaning the sampling interval for the LR EnvCF is $\Delta_x=\Delta_y=4$ m, yielding the LR EnvCF denoted as $\boldsymbol{\mathsf{F}}_{{\rm{LR}},n}$. Similarly, based on the RadioMapSeer dataset, we generate 56,000 pairs of EnvCF samples, denoted as $\left\{ {\boldsymbol{\mathsf{F}}_{{\rm{HR}},n},\boldsymbol{\mathsf{F}}_{{\rm{LR}},n}} \right\}_{n = 1}^{56000}$, and split them into training and validation sets in a $4:1$ ratio. The proposed CDiff is trained utilizing 2 Nvidia RTX-4090 GPUs, each with 24 GB of memory, and tested on a single Nvidia RTX-4090 GPU with 24 GB of memory. We employ the Adam optimizer with a learning rate of $5 \times {10^{ - 5}}$ for model parameter updates over 500,000 iterations, and the batch size is set to 16. Starting from the 5,000th iteration, we introduce the exponential moving average algorithm \cite{Ho11}, with the decay factor set to 0.9999. More parameter configurations are summarized in \tabref{ta:sys}.

\begin{figure*}[!t] 
	\centering
	\includegraphics[width=0.96\textwidth]{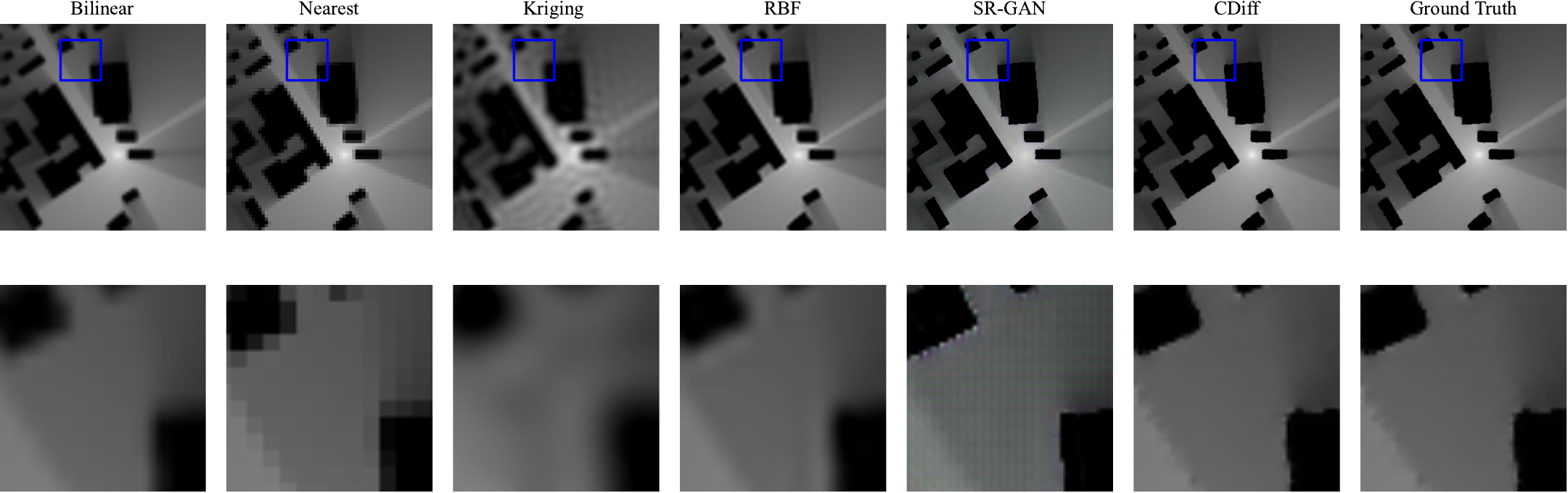}
	\captionsetup{font=footnotesize}
	\caption{Random visualizations of the HR EnvCF reconstruction results using the proposed CDiff and baselines.}
	\label{fig:srdiff}
\end{figure*}

\newcolumntype{L}{>{\hspace*{-\tabcolsep}}l}
\newcolumntype{R}{c<{\hspace*{-\tabcolsep}}}
\definecolor{lightblue}{rgb}{0.93,0.95,1.0}
\begin{table}[!b]
	\captionsetup{font=footnotesize}
	\caption{Performance comparison with baselines}\label{ta:re}
	\centering
	\setlength{\tabcolsep}{9mm}
	\ra{2}
	\scriptsize
	\scalebox{0.9}{\begin{tabular}{LccR}
			\toprule
			Method &  PSNR & SSIM & NMSE   \\
			\midrule
			\rowcolor{lightblue}
			Bilinear & 27.24 & 0.8521 & 0.0172\\
			Nearest  & 26.25 & 0.8331 & 0.0215\\
			\rowcolor{lightblue}
			Kriging  & 19.88 & 0.6725 & 0.1166\\
			RBF   & 26.99 & 0.8613 & 0.0180\\
			\rowcolor{lightblue}
			SR-GAN   & 29.75 & 0.7517 & 0.0089 \\
			CDiff  & \textbf{31.15} $\uparrow$  & \textbf{0.9280} $\uparrow$  & \textbf{0.0073} $\downarrow$ \\

			\bottomrule
		\end{tabular}
	}
\end{table}

\subsection{Experiment Results}

To comprehensively assess the effectiveness of the proposed approach, we conduct experiments on the $\times 4$ EnvVF reconstruction task, comparing its performance against several baselines, including Bilinear, Nearest, Kriging \cite{sato2017kriging}, RBF \cite{zhang2024radiomap}, and SR-GAN \cite{ledig2017photo}. Without loss of generality, three widely adopted metrics, peak signal-to-noise ratio (PSNR), structural similarity (SSIM), and normalized mean squared error (MNSE), are employed to evaluate performance. 

\tabref{ta:re} presents a quantitative analysis of the proposed CDiff and baselines on the $\times4$ EnvCF reconstruction task. It can be observed that the performance of the Kriging algorithm is relatively suboptimal. Notably, compared to the baselines, the proposed approach achieves competitive reconstruction performance, with PSNR, SSIM, and NMSE values of 31.15, 0.9280, and 0.0073, respectively. As shown in \figref{fig:srdiff}, to better illustrate the qualitative analysis, we randomly visualize the reconstruction results of EnvCF utilizing the proposed CDiff and baselines. Note that the proposed approach effectively refines both environmental information and CF, closely approximating the ground-truth EnvCF with minimal error.

\section{Conclusion}
This paper proposed a deep conditional generative learning-enabled EnvCF refinement approach that effectively refined both environmental information and CF, achieving a fourfold enhancement in granularity. Specifically, we employed variational inference to derive a surrogate objective and proposed the CDiff framework, which effectively generates HR EnvCF conditioned on LR EnvCF. Experimental results showed that the proposed approach achieved significant improvements in enhancing the granularity of EnvCF compared to the baselines.


\bibliographystyle{IEEEtran}
\bibliography{EE_AI}

\end{document}